\documentclass[twocolumn,amsmath,amssymb,floats]{jpsj3forarXiv}

\usepackage{graphicx}
\usepackage{dcolumn}
\usepackage{bm}
\usepackage{tabularx}
\usepackage{times}

\title{Spin state of Co$^{3+}$ in LaCo$_{1-x}$Rh$_{x}$O$_{3}$ investigated by structural phenomena}

\author{Shinichiro Asai$^1$, Ryuji Okazaki$^1$, Ichiro Terasaki$^1$, Yukio Yasui$^2$, Wataru Kobayashi$^3$, Akiko Nakao$^4$, Kensuke Kobayashi$^5$, Reiji Kumai$^5$, Hironori Nakao$^5$, Youichi Murakami$^5$, Naoki Igawa$^6$, Akinori Hoshikawa$^7$, Toru Ishigaki$^7$, Outi Parkkima$^8$, Maarit Karppinen$^8$, and Hisao Yamauchi$^8$}

\inst{$^1$Department of Physics, Nagoya University, Nagoya 464-8602, Japan \\
$^2$Department of Physics, Meiji University, Kawasaki 214-8571, Japan \\
$^3$Graduate School of Pure and Applied Sciences, University of Tsukuba, Ibaraki 305-8577, Japan \\
$^4$Research Center for Neutron Science and Technology, Ibaraki 319-1106, Japan \\
$^5$Condensed Matter Research Center and Photon Factory, IMSS, KEK, Ibaraki 305-0801, Japan \\
$^6$Japan Atomic Energy Agency, Ibaraki 319-1195, Japan \\
$^7$Frontier Research Center for Applied Atomic Sciences, Ibaraki University, Ibaraki 319-1106, Japan \\
$^8$Laboratory of Inorganic Chemistry, Department of Chemistry, Aalto University, Espoo FI-00076, Finland}

\abst{
Neutron diffraction for a polycrystalline sample of LaCo$_{0.8}$Rh$_{0.2}$O$_{3}$ and synchrotron x-ray diffraction for polycrystalline samples of LaCo$_{0.9}$Rh$_{0.1}$O$_{3}$ and LaCo$_{0.8}$Rh$_{0.2}$O$_{3}$ have been carried out in order to investigate the structural properties related with the spin state of Co$^{3+}$ ions.
We have found that the values of the Co(Rh)-O bond lengths in the Co(Rh)O$_{6}$ octahedron of LaCo$_{0.8}$Rh$_{0.2}$O$_{3}$ are nearly identical at 10 K.
The lattice volume for the Rh$^{3+}$ substituted samples decreases with the thermal expansion coefficient similar to that of LaCoO$_{3}$ from room temperature, and ceases to decrease around 70 K.
These experimental results favor a mixed state consisting of the high-spin-state and low-spin-state Co$^{3+}$ ions, and suggest that the high-spin-state Co$^{3+}$ ions are thermally excited in addition to those pinned by the substituted Rh$^{3+}$ ions.
}

\kword{LaCo$_{1-x}$Rh$_{x}$O$_{3}$, X-ray scattering, Neutron diffraction, Spin crossover, Spin state}
\begin{document}
\maketitle

\section{Introduction}
The transition-metal oxides containing a Co$^{3+}$ ion surrounded by six O$^{2-}$ ions octahedrally have been extensively studied because the Co$^{3+}$ ion can take three spin states~\cite{JAP.36.879}.
The fivefold degenerate 3\textit{d} orbitals in the Co$^{3+}$ ion are split into the doubly degenerate \textit{e$_{g}$} orbitals in the upper energy level and the triply degenerate \textit{t$_{2g}$} orbitals in the lower energy level owing to the Coulomb interaction between the \textit{d} electrons and O$^{2-}$ ions.
The Co$^{3+}$ ions take the low-spin state (LS: \textit{t}$_{2g}^{6}$; S = 0), when the crystal field splitting is larger than the Hund coupling.
In contrast, they take the high-spin state (HS: \textit{e}$_{g}^{2}$\textit{t}$_{2g}^{4}$; S = 2) for the opposite condition.
The atomic multiplet calculations suggest that the intermediate-spin state (IS: \textit{e}$_{g}^{1}$\textit{t}$_{2g}^{5}$; S = 1) can be realized when the hybridization between the Co \textit{e}$_{g}$ orbitals and O 2\textit{p} orbitals is taken into account~\cite{PhysRevB.51.11501}.

The spin state of the Co$^{3+}$ ions in the perovskite oxide LaCoO$_{3}$ has been long discussed theoretically and experimentally because it changes with temperature~\cite{Physica.30.1600}, magnetic field~\cite{JPSJ.78.093702}, or pressure~\cite{PRB.67.140401}, known as the spin-state crossover.
NMR~\cite{JPSJ.64.3967} and neutron scattering measurements~\cite{PRB.40.10982} for LaCoO$_{3}$ show that the ground state of the Co$^{3+}$ ions is the low-spin state. 
The magnetic susceptibility of LaCoO$_{3}$ increases with increasing temperature up to 90 K except for the contribution owing to the lattice defects~\cite{PhysRevB.79.174410},
which indicates that the magnetically excited state is thermally activated and the spin-state crossover takes place with temperature~\cite{JPSJ.64.3967}. 
Anomalous expansion of the lattice volume for LaCoO$_{3}$ is observed via the powder neutron diffraction, which implies that the Co$^{3+}$ ion with a large ionic radius is thermally excited~\cite{PRB.66.094408}.

The excited state of LaCoO$_{3}$ at room temperature is controversial. 
The magnetic susceptibility of LaCoO$_{3}$ at room temperature shows Curie-Weiss-like behavior with an effective magnetic moment $\mu_{\rm eff}$ of 3.1 $\mu_{\rm B}$,
which is smaller than that expected for the HS Co$^{3+}$ ions (4.9 $\mu_{\rm B}$)~\cite{Physica.30.1600}.
An NaCl-type spin-state order consisting of the high-spin and low-spin states of the Co$^{3+}$ ions accompanied by the displacements of the O$^{2-}$ ions was proposed as an excited state in LaCoO$_{3}$~\cite{PR.155.932},
but no evidence for such a spin-state order is observed until now.
A dynamically ordered state consisting of the high-spin and low-spin states (HS-LS model) is alternatively suggested from the soft x-ray-absorption spectroscopy~\cite{PRL.97.176405} and heat capacity combined with the magnetic susceptibility~\cite{PRB.71.024418,PRB.67.144424}.
The calculations using the local-density approximation (LDA+\textit{U}) suggest that the excited state of the Co$^{3+}$ ions in LaCoO$_{3}$ is the intermediate-spin state (IS model)~\cite{PRB.54.5309}.
They propose the antiferro orbital ordering owing to the Jahn-Teller distortion of the CoO$_{6}$ octahedra, which is suggested by the synchrotron x-ray diffraction~\cite{PRB.67.224423}.
The IS model is also supported by the x-ray photoemission spectroscopy~\cite{PRB.55.4257} and infrared spectroscopy~\cite{PhysRevB.55.R8666}.

We have studied Rh substitution effects in LaCoO$_{3}$ in order to clarify the spin state of the Co$^{3+}$ ions.
Partial substitution of Rh$^{3+}$ ($t_{2g}^{6}$; S = 0) for Co$^{3+}$ in LaCoO$_{3}$ suppresses the spin-state crossover and a Curie-Weiss like susceptibility develops down to low temperatures for $x > 0.04$ in LaCo$_{1-x}$Rh$_{x}$O$_{3}$~\cite{PRB.67.144424}.
In contrast to the significant change of the magnetic susceptibility, the electrical resistivity of LaCo$_{1-x}$Rh$_{x}$O$_{3}$ is non-metallic, as is similar to that of LaCoO$_{3}$~\cite{JSSC.183.1388}.
First principle band calculations indicate that the unfilled 4\textit{d} shell and covalency of the Rh$^{3+}$ ion stabilizes the HS Co$^{3+}$ ions~\cite{PhysRevB.85.134401}.
We found that the effective magnetic moment of LaCo$_{1-x}$Rh$_{x}$O$_{3}$ evaluated at room temperature is independent of Rh content $x$ for $0 \leq x \leq 0.5$,
which suggests that the substituted Rh$^{3+}$ ion behaves as if it were preferentially substituted for a LS Co$^{3+}$ ion~\cite{JPSJ.80.104705}.
This suggestion is supported by the non-monotonic variation in the lattice volume for LaCo$_{1-x}$Rh$_{x}$O$_{3}$.
Next, we investigated the Ga$^{3+}$ substitution effects for Co$^{3+}$ in LaCo$_{0.8}$Rh$_{0.2}$O$_{3}$ and found that the substituted Ga$^{3+}$ ion behaves as if it were preferentially substituted for a HS Co$^{3+}$ ion~\cite{PhysRevB.86.014421}.
We further found a ferromagnetic ordering of LaCo$_{1-x}$Rh$_{x}$O$_{3}$ below 15 K in the range of $0.1 \leq x \leq 0.4$~\cite{JPSJ.80.104705}.
The ferromagnetic ordering in the insulating phase is driven only by Co$^{3+}$ ions, and is different from the metallic ferromagnetic state of La$_{1-x}$Sr$_{x}$CoO$_{3}$ occurring in the mixed valence of Co$^{3+}$ and Co$^{4+}$~\cite{Physica.19.120}.

As discussed above, the spin-state crossover affects not only the magnetic properties but also the structural properties.
The detailed structural analysis is expected to reveal the spin state of Co$^{3+}$ through the distortion of the CoO$_{6}$ octahedron.
In this paper, we investigate the structural properties of LaCo$_{0.9}$Rh$_{0.1}$O$_{3}$ and LaCo$_{0.8}$Rh$_{0.2}$O$_{3}$ related with the spin-state crossover by using the synchrotron x-ray diffraction.
We also discuss the spin state of Co$^{3+}$ in LaCo$_{0.8}$Rh$_{0.2}$O$_{3}$ via the shape and size of the Co(Rh)O$_{6}$ octahedron determined by the neutron diffraction which can precisely determine the position of the O$^{2-}$ ions.
We can associate the distortion of the CoO$_{6}$ octahedron of LaCo$_{0.8}$Rh$_{0.2}$O$_{3}$ with the spin state of Co$^{3+}$ ions because its space group is \textit{Pnma} where oxygen ions occupy two inequivalent sites.
This makes a remarkable contrast to LaCoO$_{3}$, where all the oxygen sites are equivalent.
We show that the HS-LS model is more likely, and that the HS Co$^{3+}$ ions are thermally excited in addition to those stabilized by Rh$^{3+}$ substitution.

\section{Experiments}
Polycrystalline samples of LaCo$_{1-x}$Rh$_{x}$O$_{3}$ (\textit{x} = 0.1 and 0.2) were prepared by a standard solid-state reaction method.
A mixture of La$_{2}$O$_{3}$ (3\textit{N}), Co$_{3}$O$_{4}$ (3\textit{N}), and Rh$_{2}$O$_{3}$ (3\textit{N}) with stoichiometric molar ratio was ground, and
calcined for 24 h at 1000 $^{\circ}$C in air.
The calcined powder was ground, pressed into a pellet, and sintered for 48 h at 1200 $^{\circ}$C in air.
These samples were characterized by x-ray diffraction measured with Rigaku 4037V (Cu K$\alpha$ radiation), and no impurity phases were detected.
The oxygen content of LaCo$_{0.8}$Rh$_{0.2}$O$_{3}$ was determined by the thermogravimetric H$_{2}$ reduction analysis.
The synchrotron powder x-ray diffraction for LaCo$_{0.9}$Rh$_{0.1}$O$_{3}$ and LaCo$_{0.8}$Rh$_{0.2}$O$_{3}$ was carried out at BL-8A and BL-8B, Photon Factory, KEK, Japan.
Wavelengths of 0.7755 \AA \space and 0.6884 \AA \space were selected for the measurements from 100 to 300 K (BL-8A) and from 30 to 100 K(BL-8B), respectively.
The diffraction patterns were analyzed by the Rietveld refinement using RIETAN-FP~\cite{rietanfp}.
Neutron powder diffraction for LaCo$_{0.8}$Rh$_{0.2}$O$_{3}$ was carried out at 10 K at iMATERIA, MLF, J-PARC, Japan.
The obtained diffraction pattern was analyzed by Z-rietveld 0.9.37.2~\cite{zrietveld1,zrietveld2}.

\section{Results and discussion}
Thermogravimetric reduction experiment yielded the oxygen content ${3 -\delta }$ of LaCo$_{0.8}$Rh$_{0.2}$O$_{3 -\delta }$ sample at 2.95 $\pm$ 0.01.
This value is essentially the same as those obtained for the La$_{0.8}$Sr$_{0.2}$Co$_{1-x}$Rh$_{x}$O$_{3-\delta }$ system with the same analysis technique~\cite{PhysRevB.83.094405},
and indicates only a tiny deviation from the stoichiometric oxygen content value.
In the following we denote the two samples with their stoichiometric oxygen-content value.

\begin{figure}
\begin{center}
\includegraphics[width=70.0mm,clip]{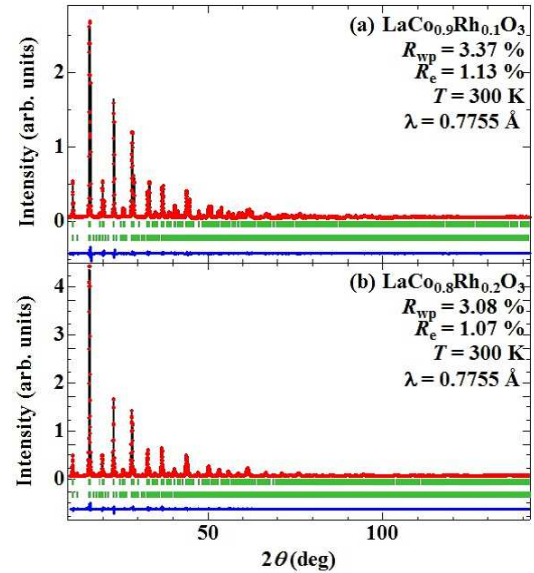}\\
\caption{(Color online) The synchrotron powder x-ray diffraction patterns for (a) LaCo$_{0.9}$Rh$_{0.1}$O$_{3}$ and (b) LaCo$_{0.8}$Rh$_{0.2}$O$_{3}$ at 300 K.
The solid circle and solid line above the vertical bars represent the observed pattern and calculated pattern, respectively.
Upper and lower vertical bars in the middle show the position of the Bragg peaks for the \textit{R-3c} and \textit{Pnma} phases, respectively.
The solid line below the vertical bars shows the difference curve.}
\end{center}
\end{figure}

Figures 1(a) and 1(b) show the synchrotron powder x-ray diffraction patterns for LaCo$_{0.9}$Rh$_{0.1}$O$_{3}$ and LaCo$_{0.8}$Rh$_{0.2}$O$_{3}$ at 300 K, respectively.
We have analyzed these profiles as a mixture of two phases, rhombohedral \textit{R-3c} and orthorhombic \textit{Pnma}.
It should be noted that LaCo$_{0.9}$Rh$_{0.1}$O$_{3}$ and LaCo$_{0.8}$Rh$_{0.2}$O$_{3}$ are located near the boundary between \textit{R-3c} and \textit{Pnma} phases in the phase diagram of LaCo$_{1-x}$Rh$_{x}$O$_{3}$ at room temperature~\cite{JPSJ.80.104705}.
As shown in Fig. 1, the refinements give satisfactory reliable factors.

\begin{figure}[!htb]
\begin{center}
\includegraphics[width=70.0mm,clip]{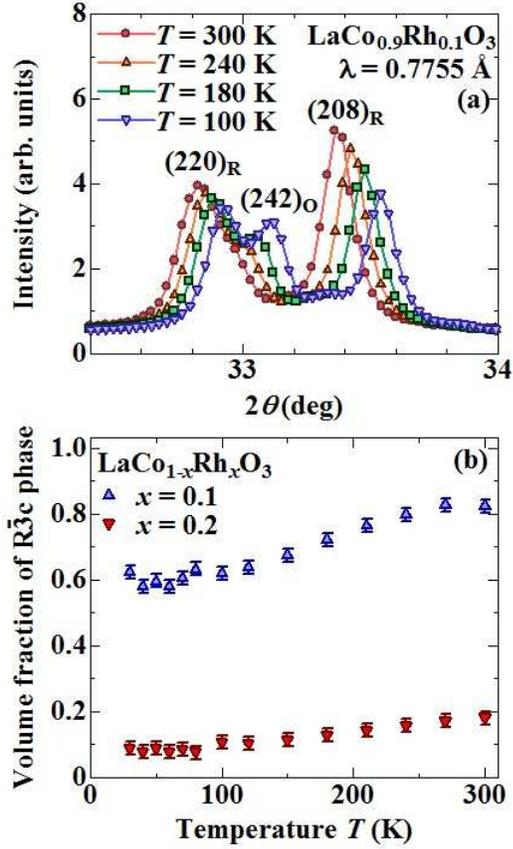}\\
\caption{(Color online) (a)The x-ray diffraction patterns for LaCo$_{0.9}$Rh$_{0.1}$O$_{3}$ at 100, 180, 240, and 300 K.
(220)$_{\rm R}$ and (208)$_{\rm R}$ represent the (220) and (208) peaks of the \textit{R-3c} phase, and (242)$_{\rm O}$ does the (242) peak of the \textit{Pnma} phase.
(b)The temperature dependence of the volume fraction of the \textit{R-3c} phase of LaCo$_{0.9}$Rh$_{0.1}$O$_{3}$ and LaCo$_{0.8}$Rh$_{0.2}$O$_{3}$.}
\end{center}
\end{figure}
Figure 2(a) shows the x-ray diffraction patterns for LaCo$_{0.9}$Rh$_{0.1}$O$_{3}$ at 100, 180, 240, and 300 K.
The profile at 300 K shows two peaks which are indexed as \textit{R-3c}.
Another peak which cannot be indexed as \textit{R-3c} develops near 33 deg with decreasing temperature.
This peak is indexed as (242) in the \textit{Pnma} phase, meaning that the volume fraction of the two phases change with temperature.
Thus we analyzed the volume fraction at every temperature by fitting the diffraction patterns with the two phase mixture.
The temperature dependence of the volume fraction of the \textit{R-3c} phase for LaCo$_{0.9}$Rh$_{0.1}$O$_{3}$ and LaCo$_{0.8}$Rh$_{0.2}$O$_{3}$ is shown in Fig. 2(b).
The volume fraction of the \textit{R-3c} phase decreases with increasing \textit{x} or with decreasing temperature,
implying that the crystal structure inhomogeneously changes from the \textit{R-3c} to \textit{Pnma} phase as a function of temperature and Rh content \textit{x} as a mixed phase.
Similar structural change and phase mixture are also observed in La$_{1-x}$Ca$_{x}$CoO$_{3}$~\cite{PhysRevB.69.054401}.
Ca$^{2+}$ substitution for La$^{3+}$ and Rh$^{3+}$ substitution for Co$^{3+}$ similarly decreases the tolerance factor of LaCoO$_{3}$, which suggests that the lattice distortion caused by the substituted Rh$^{3+}$ ions is the origin of the structural changes in LaCo$_{1-x}$Rh$_{x}$O$_{3}$.

\begin{figure}
\begin{center}
\includegraphics[width=70.0mm,clip]{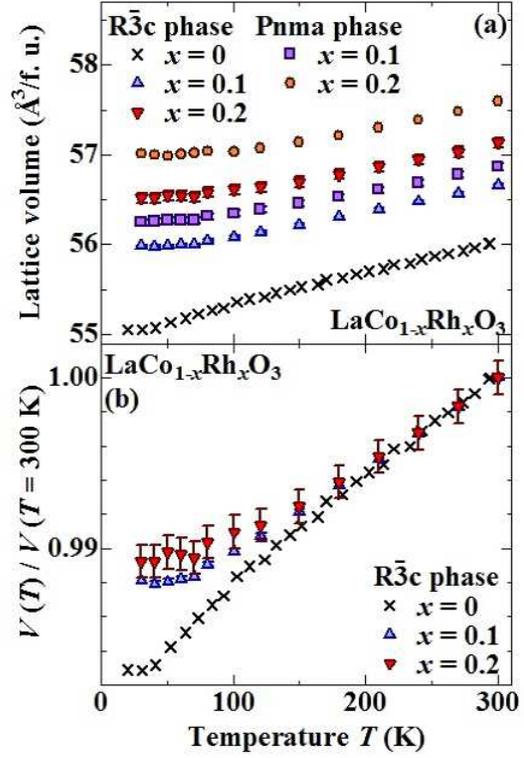}\\
\caption{
(Color online) (a)The lattice volume of LaCo$_{0.9}$Rh$_{0.1}$O$_{3}$ and LaCo$_{0.8}$Rh$_{0.2}$O$_{3}$ as a function of temperature.
The temperature dependence of the lattice volume for LaCoO$_{3}$ taken from Ref. 28 is also plotted.
(b)The temperature dependence of the lattice volume normalized by \textit{V}(300 K) 
for \textit{R-3c} phase of LaCo$_{1-x}$Rh$_{x}$O$_{3}$ (\textit{x} = 0 $\sim$  0.2).

}
\end{center}
\end{figure}

Figure 3(a) shows the lattice volume of LaCo$_{0.9}$Rh$_{0.1}$O$_{3}$ and LaCo$_{0.8}$Rh$_{0.2}$O$_{3}$ as a function of temperature 
together with the temperature dependence of the lattice volume for LaCoO$_{3}$ reported by K. Asai \textit{et al}~\cite{JPSJ.67.290,methods1}.
The lattice volume for the two phases increase with increasing Rh content \textit{x}, which is consistent with the previous study~\cite{JPSJ.80.104705}.
The lattice volume of the \textit{Pnma} phase is larger than that of \textit{R-3c} phase in the same composition.
Figure 3(b) shows the temperature dependence of the lattice volume normalized by the value at 300 K for \textit{R-3c} phase of LaCo$_{1-x}$Rh$_{x}$O$_{3}$ (\textit{x} = 0 $\sim$  0.2).
It should be noted that the lattice volume of LaCoO$_{3}$ continues to decrease with decreasing temperature down to 40 K and that the thermal expansion coefficient of LaCoO$_{3}$ enhances owing to the spin-state crossover.
The lattice volume of LaCo$_{0.9}$Rh$_{0.1}$O$_{3}$ and LaCo$_{0.8}$Rh$_{0.2}$O$_{3}$ decreases with the thermal expansion coefficient similar to that of LaCoO$_{3}$ from room temperature, which indicates that the HS Co$^{3+}$ ions are thermally excited in Rh$^{3+}$ substituted samples.
The lattice volume ceases to decrease around 70 K in Rh$^{3+}$ substituted samples, which shows that Rh$^{3+}$ substitution suppresses the spin-state crossover below 70 K.

\begin{figure}
\begin{center}
\includegraphics[width=70.0mm,clip]{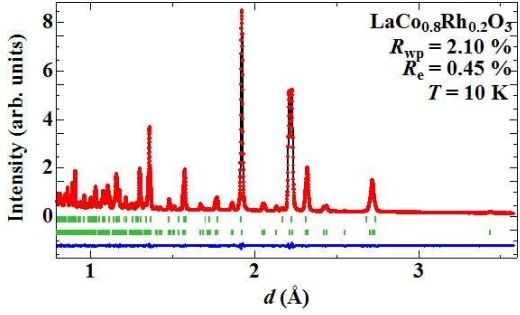}\\
\caption{(Color online) The neutron diffraction pattern for LaCo$_{0.8}$Rh$_{0.2}$O$_{3}$ at 10 K.
The solid circle and solid line above the vertical bars represent the observed pattern, calculated pattern, respectively.
Upper and lower vertical bars in the middle show the position of the Bragg peaks for the \textit{R-3c} and \textit{Pnma} phases, respectively.
The solid line below the vertical bars shows the difference curve.}
\end{center}
\end{figure}

\begin{table}
\begin{center}
\label{table2}
\caption{The structural parameters of LaCo$_{0.8}$Rh$_{0.2}$O$_{3}$ from the neutron diffraction at 10 K.}
\smallskip
\small
\begin{tabular}{cccccc}
\hline
Atom & Site & \textit{x} & \textit{y} & \textit{z} & \textit{B} (\AA $^{2}$) \\ \hline
La & 4c & 0.4753(2) & 0.25 & 0.5039(1) & 0.07(5)\\
Co(Rh) & 4b & 0 & 0 & 0.5 & 0.18(1)\\
O1 & 4c & 0.004(1) & 0.25 & 0.4336(5) & 0.35(1)\\
O2 & 8d & 0.2232(2) & 0.0353(1) & 0.7761(1) & 0.35(1)\\
\hline
\end{tabular}
\end{center}
\end{table}

\begin{table}
\begin{center}
\label{table3}
\caption{The evaluated bond lengths and bond angles of LaCo$_{0.8}$Rh$_{0.2}$O$_{3}$. M = Co(Rh).}
\smallskip
\small
\begin{tabular}{cccc}
\hline
Bond length(\AA) & & Bond angle (deg.) & \\ \hline
M-O1 $\times $ 2 & 1.9592(6) & M-O1-M $\times $ 2 & 158.57(15)\\
M-O2 $\times $ 2 & 1.9540(9) & M-O2-M $\times $ 4 & 159.99(2) \\
M-O2 $\times $ 2 & 1.9545(6) & O1-M-O2 $\times $ 2 & 90.00(3)\\
          &          & O1-M-O2 $\times $ 2 & 90.68(2)\\
          &          & O2-M-O2 $\times $ 2 & 91.75(1)\\
\hline
\end{tabular}
\end{center}
\end{table}

Figure 4 shows the neutron diffraction pattern for LaCo$_{0.8}$Rh$_{0.2}$O$_{3}$ at 10 K.
We evaluate 3 $\%$ \textit{R-3c} phase from the analysis, which is consistent with the experimental results of the synchrotron x-ray diffraction as was discussed in Fig. 2.
We will not discuss the crystal structure for the \textit{R-3c} phase because the small volume fraction makes it difficult to evaluate the structural parameters of the \textit{R-3c} phase.
The lattice parameters \textit{a} = 5.4125(6) \AA, \textit{b} = 7.7002(6) \AA, and \textit{c} = 5.4739(2) \AA \space for the \textit{Pnma} phase are evaluated from the neutron diffraction shown in Fig. 4.
Tables I and I\hspace{-.1em}I list the structural parameters of LaCo$_{0.8}$Rh$_{0.2}$O$_{3}$ and the bond lengths and bond angles of LaCo$_{0.8}$Rh$_{0.2}$O$_{3}$, respectively.

Here we discuss the relation between the spin state of the Co$^{3+}$ ion and the shape of the Co(Rh)O$_{6}$ octahedron.
The HS, IS, and LS Co$^{3+}$ ion have 2, 1, and 0 electrons in \textit{e$_{g}$} orbitals, respectively.
The IS Co$^{3+}$ ion has the Jahn-Teller instability at the \textit{e$_{g}$} orbitals.
Thus, when the CoO$_{6}$ octahedron contains the IS Co$^{3+}$ ions, the bond lengths will have anisotropic values because of the Jahn-Teller distortion.  
On the other hand, when the CoO$_{6}$ octahedron contains the HS or LS Co$^{3+}$ ions, the bond lengths are expected to be identical.
The Co(Rh)O$_{6}$ octahedron can be regarded as an average of the isotropic RhO$_{6}$ octahedra and CoO$_{6}$ octahedra, and we simply assume that Rh substitution dilutes the distortion of CoO$_{6}$ octahedron.
If the anti-ferro orbital ordering as discussed in the IS model~\cite{PRB.54.5309} also occurred in LaCo$_{0.8}$Rh$_{0.2}$O$_{3}$, we could evaluate the diluted Jahn-Teller distortion from our structural analysis.
The difference in the Co-O bond lengths $\Delta r$ is estimated to be 0.15 \AA \space for the CoO$_{6}$ octahedron which contains the IS Co$^{3+}$ ion reported for Sr$_{3}$YCo$_{4}$O$_{10.5+\delta }$~\cite{PhysRevB.80.024409}. 
If the IS model were valid for LaCo$_{0.8}$Rh$_{0.2}$O$_{3}$, $\Delta r$ of LaCo$_{0.8}$Rh$_{0.2}$O$_{3}$ would be expected to be 0.12 \AA \space which is 20 times larger than the observed $\Delta r$ (0.005 \AA ).
Thus we conclude from a structural point of view that the content of the IS Co$^{3+}$ ions is at most less than 4 \%.
Obviously this amount cannot explain the magnetization of LaCo$_{0.8}$Rh$_{0.2}$O$_{3}$.
The HS Co$^{3+}$ ions mixed with the LS Co$^{3+}$ ions can naturally explain the magnetization and the shape of the Co(Rh)O$_{6}$.
It should be noted that no structural phase transition accompanied by the Jahn-Teller distortion is observed for LaCo$_{0.8}$Rh$_{0.2}$O$_{3}$ in our structural analysis, which suggests that the Co$^{3+}$ ions in LaCo$_{0.8}$Rh$_{0.2}$O$_{3}$ have no orbital degree of the freedom.
These experimental results favor the HS-LS model rather than the IS model, as is consistent with our previous study~\cite{JPSJ.80.104705,PhysRevB.86.014421}.

Next, we discuss the volume $V_{\rm obs}$ of the Co(Rh)O$_{6}$ octahedron in LaCo$_{0.8}$Rh$_{0.2}$O$_{3}$, which is determined to be 9.97 \AA $^{3}$ by using the structural parameters given in Table I\hspace{-.1em}I.
$V_{\rm obs}$ can be regarded as the average value of the octahedra which have the HS Co$^{3+}$ ions, the LS Co$^{3+}$ ions, and the Rh$^{3+}$ ions.
Thus let us calculate the volume by using the following equation,
\begin{equation}
V_{\rm cal}=0.8 V_{\rm LCO} + 0.2 V_{\rm LRO}.
\end{equation}
$V_{\rm LCO}$ and $V_{\rm LRO}$ represent the volume of the CoO$_{6}$ octahedron in LaCoO$_{3}$ and that of the RhO$_{6}$ octahedron in LaRhO$_{3}$, respectively.
$V_{\rm LCO}$ is evaluated from the structural parameters reported in Ref. 9.
$V_{\rm LRO}$ is set to be 11.54 \AA $^{3}$ estimated from the x-ray diffraction at 294 K~\cite{CGD61361} (no structure data are available for LaRhO$_{3}$ at low temperature).
At first, we calculate $V_{\rm cal}$ for the LS Co$^{3+}$ ions.
$V_{\rm LCO}$  (= 9.51 \AA $^{3}$) is taken from the structural parameters at 10 K, and then we get $V_{\rm cal}$ = 9.92 \AA $^{3}$, which is substantially smaller than $V_{\rm obs}$.
This indicates that there are other spin-state Co$^{3+}$ ions in LaCo$_{0.8}$Rh$_{0.2}$O$_{3}$, which is consistent with our previous study~\cite{JPSJ.80.104705,PhysRevB.86.014421}.
$V_{\rm obs}$ is close to $V_{\rm cal}$ when the structural parameters of LaCoO$_{3}$ at 150 K are used for calculation ($V_{\rm LCO}$ = 9.58  \AA $^{3}$).
The previous study~\cite{PRB.67.144424} suggests that the content of the HS Co$^{3+}$ ions in LaCoO$_{3}$ at 150 K is 80 \% of that at 300 K.
Taking the dilution effect by the substituted Rh$^{3+}$ ions into account, we estimate the content of the HS Co$^{3+}$ ions for LaCo$_{0.8}$Rh$_{0.2}$O$_{3}$ at 10 K to be 64 \% of that for LaCoO$_{3}$ at 300 K.
According to our previous study~\cite{JPSJ.80.104705}, LaCo$_{0.8}$Rh$_{0.2}$O$_{3}$ has almost the same amount of the HS Co$^{3+}$ as LaCoO$_{3}$ does at room temperature.
The difference between the contents of the HS Co$^{3+}$ ions between 10 and 300 K suggests that the HS Co$^{3+}$ ions are thermally excited in addition to those stabilized by Rh$^{3+}$ substitution at low temperature.

\begin{figure}
\begin{center}
\includegraphics[width=70.0mm,clip]{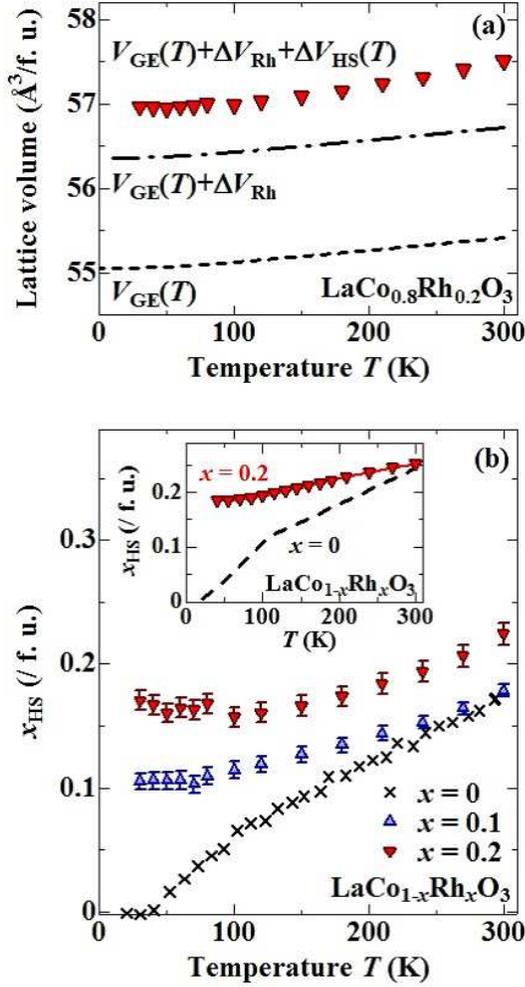}\\
\caption{
(Color online) (a)The lattice volume of LaCo$_{0.8}$Rh$_{0.2}$O$_{3}$ for a mixture of two phases as a function of temperature.
The dotted and dot-dashed lines represent $V_{\rm{GE}}(T)$ and $V_{\rm{GE}}(T)$ + $\Delta V_{\rm{Rh}}$ evaluated for LaCo$_{0.8}$Rh$_{0.2}$O$_{3}$, respectively (see text).
(b)The content of the HS Co$^{3+}$ ions for LaCo$_{1-x}$Rh$_{x}$O$_{3}$ (\textit{x} = 0 $\sim$  0.2) as a function of temperature.
The inset shows the content of the HS Co$^{3+}$ ions obtained from the susceptibility as a function of temperature~\cite{PhysRevB.86.014421}.
}
\end{center}
\end{figure}
In order to discuss the spin-state crossover of LaCo$_{1-x}$Rh$_{x}$O$_{3}$ more quantitatively, we evaluate the contribution to the spin-state crossover for the lattice expansion by subtracting the other contributions from the experimental results.
At first, we calculate the temperature dependence of the lattice volume for LaCoO$_{3}$ expected from the phonon contribution by using the Gr\"{u}neisen-Einstein formula given by
\begin{equation}
V_{\rm{GE}}(T)=V(\rm{0~K})\{ 1+\frac{\alpha_{V}\textit{T}_{E}}{2}(\coth(\frac{\textit{T}_{E}}{2\textit{T}})-1) \},
\end{equation}
where $T_{\rm E}$ and $\alpha_{\rm V}$ represent the Einstein temperature and the thermal expansion coefficient for $\textit{T} \gg T_{\rm E}$, respectively.
The value of $V$(0~K) (= 55.06 \AA $^{3}$) is evaluated from extrapolating the lattice volume of LaCoO$_{3}$ shown in Fig. 3(a) towards 0 K.
We do not employ $T_{\rm E}$ and $\alpha_{\rm V}$ for LaCoO$_{3}$ because the spin-state crossover affects the temperature dependence of the lattice volume.
Instead, we use $T_{\rm E}$ and $\alpha_{\rm V}$ for La$_{0.7}$Sr$_{0.3}$CoO$_{3}$~\cite{Europhys45399}, which are estimated to be 142 K and 2.76 $\times $ 10$^{-5}$ K$^{-1}$, respectively.
Next, we assume that the lattice also expands through the increase of an average ionic radius for Co(Rh) sites (the B sites) as reported in Ref. 9.
Then we can obtain the additional lattice expansion $\Delta V_{\rm{B}}$ due to the HS Co$^{3+}$ and Rh$^{3+}$ ions as
\begin{equation}
\Delta V_{\rm{B}}=x_{\rm{B}}(r_{\rm{B}}-r_{\rm{LS}})\frac{\partial V}{\partial r_{\rm{B}}},
\end{equation} 
where $x_{\rm{B}}$ and $r_{\rm{B}}$ represent the content and ionic radius of B ions, respectively.
As for the B ion, we abbreviate HS Co$^{3+}$, LS Co$^{3+}$, and Rh$^{3+}$ as HS, LS, and Rh, respectively.
Finally we arrive at the expression for the lattice volume of Rh$^{3+}$ substituted samples given by
\begin{equation}
V(T)=V_{\rm{GE}}(T)+\Delta V_{\rm{Rh}}+\Delta V_{\rm{HS}}(T).
\end{equation} 
$V(T)$ for a mixture of two phases are evaluated as an average from the lattice volume and volume fraction of two phases shown above.
According to the analysis in Ref. 9, We evaluate $ \partial V / \partial r_{\rm{B}}$ to be 54.38 \AA $^{2}$ from the lattice volume of LaBO$_{3}$ as a function of the ionic radius for B ions~\cite{fitting}.
The values of $r_{\rm{B}}$ for B = LS, HS, and Rh are set to be 0.545, 0.61, and 0.665 \AA, respectively~\cite{Shannon}.
Now we can calculate $\Delta V_{\rm{Rh}}$ by using $x_{\rm{Rh}}$ taken from the chemical formula of the samples.
The contributions of $V_{\rm{GE}}(T)$ and $\Delta V_{\rm{Rh}}$ together with $V(T)$ for LaCo$_{0.8}$Rh$_{0.2}$O$_{3}$ are shown in Fig. 5(a).
We can evaluate the $\Delta V_{\rm{HS}}(T)$ by subtracting $V_{\rm{GE}}(T)$ + $\Delta V_{\rm{Rh}}$ from $V(T)$.
Figure 5(b) shows the HS Co$^{3+}$ content $x_{\rm{HS}}$ evaluated from $\Delta V_{\rm{HS}}(T)$ for LaCo$_{1-x}$Rh$_{x}$O$_{3}$ (\textit{x} = 0 $\sim$  0.2) plotted as a function of temperature.
$x_{\rm{HS}}$ at room temperature is around 0.2 and almost independent of Rh$^{3+}$ content,
which indicates that Rh$^{3+}$ substitution does not change the HS Co$^{3+}$ content at room temperature.
While $x_{\rm{HS}}$ of LaCoO$_{3}$ decreases to zero with decreasing temperature,
$x_{\rm{HS}}$ of Rh$^{3+}$ substituted samples decreases from room temperature, but ceases to decrease below 70 K.
We emphasize that $x_{\rm{HS}}$ evaluated from the lattice volume is quantitatively consistent with that obtained from the susceptibility as was reported in our previous study~\cite{PhysRevB.86.014421}, as shown in the inset of Fig. 5(b).
We expect that the gradual change of the spin state of Co$^{3+}$ ions above 100 K is due to the thermal excitation of the HS Co$^{3+}$ ions disturbed by the repulsion between the HS Co$^{3+}$ ions as discussed in LaCoO$_{3}$~\cite{PRB.67.144424,PRB.71.024418}.
Further investigation is needed in order to clarify the relation between the thermally excited HS Co$^{3+}$ ions and those pinned by Rh$^{3+}$ substitution.

\section{Summary}
We have carried out the neutron powder diffraction for LaCo$_{0.8}$Rh$_{0.2}$O$_{3}$ and synchrotron powder x-ray diffraction for LaCo$_{0.9}$Rh$_{0.1}$O$_{3}$ and LaCo$_{0.8}$Rh$_{0.2}$O$_{3}$.
The thermal expansion coefficient is almost independent of Rh content, which suggests that the high-spin-state Co$^{3+}$ ions are thermally excited above 100 K in the Rh$^{3+}$ substituted samples.
The CoO$_{6}$ octahedra of LaCo$_{0.8}$Rh$_{0.2}$O$_{3}$ determined by the neutron diffraction are found to be isotropic, which supports the HS-LS model.
The volume of the Co(Rh)O$_{6}$ octahedra of LaCo$_{0.8}$Rh$_{0.2}$O$_{3}$ (9.97 \AA $^{3}$) is larger than that expected from the mixture of the low-spin-state Co$^{3+}$ and Rh$^{3+}$ ions.
The calculation using the temperature dependence of the lattice volume of LaCoO$_{3}$ indicates that the content of the high-spin-state Co$^{3+}$ ions for LaCo$_{0.8}$Rh$_{0.2}$O$_{3}$ at 10 K is 64 \% of that for LaCoO$_{3}$ at 300 K.
We evaluate the content of the high-spin-state Co$^{3+}$ ions for LaCo$_{1-x}$Rh$_{x}$O$_{3}$ (\textit{x} = 0 $\sim$  0.2) as a function of temperature from the temperature dependence of the lattice volume.
The content of the high-spin-state Co$^{3+}$ ions of Rh$^{3+}$ substituted samples decreases with decreasing temperature as that of LaCoO$_{3}$ does at room temperature, and ceases to decrease below 70 K.
These experimental results suggest that the high-spin-state Co$^{3+}$ ions are thermally excited in addition to those stabilized by Rh$^{3+}$ substitution.

\section*{Acknowledgments}
This work is partially supported by Program for Leading Graduate Schools, Japan Society for Promotion of Science and by Advanced Low Carbon Research and Development Program, Japan Science and Technology Agency.
The x-ray diffraction was performed under the approval of the Photon Factory Advisory Committee (No. 2009S2-008).
The neutron diffraction was performed under the approval of the Proposal Review Committee of J-PARC MLF (No. 2012A0125).


\begin{thebibliography}{99}
\bibitem{methods1}
The results which use different wavelengths are combined by converting the value of lattice volume measured by 0.7755 \AA \space in such a way that the converted values of the lattice volume at 100 and 300 K are same with those measured by 0.6884 \AA .
\bibitem{fitting}
LaCuO$_{3}$ is excluded from the fitting in Ref. 9.
In addition, we exclude LaMnO$_{3}$ and LaCoO$_{3}$ from the fitting because the simple definition for the effective ionic radius is not valid owing to the Jahn-Teller effect and spin-state crossover, respectively.
\end{thebibliography}
\end{document}